\documentstyle[preprint,aps]{revtex}
\begin{document}
\title{Numerical Renormalization Group for Finite Fermi Systems}
\author{Taku Tokuyasu}
\address{Department of Physics, Boston University}
\author{Michael Kamal}
\address{Department of Physics, Harvard University}
\author{Ganpathy Murthy}
\address{Department of Physics, Boston University}
\maketitle

\widetext
\begin{abstract}
A numerical
renormalization group technique based on Wilson's momentum shell method
is presented for interacting, finite fermi systems. Results for small fullerene
analogs  show that the method is quite accurate to moderate values of
$U$, and qualitatively better than other approximation methods. The
method  has potential applicability far beyond the fullerenes.
\end{abstract}

\narrowtext

The discovery of superconductivity in the high-$T_c$
compounds\cite{hightc}  and now
$C_{60}$\  \cite{c60} has spurred a remarkable surge of interest
in the theory of interacting fermi systems.
Despite the  complexity of such systems,
condensed matter theorists have had
many successes such as the BCS theory of superconductivity\cite{bcs}
in describing their behavior.
These successes have
been based on
the phenomenological framework of Landau fermi liquid theory\cite{flt},
augmented by physically motivated approximations.
More recently, attention has focussed on phases
which  challenge the foundations of fermi
liquid theory itself\cite{mflt}.
Our conception of interacting
fermi systems has broadened considerably due to these efforts,
and it is apparent that we need a way
to describe these systems in a unified manner.

As pioneered by Anderson and Yuval\cite{yuval} and developed by many
authors\cite{hertz,solyom,whiterg1,benef,shankar},
the renormalization group (RG) method\cite{wils}, which was
developed to describe phase transitions in classical systems,
provides a natural framework for the study of fermionic systems.
In particular, the RG provides a powerful tool for bridging
the gap between the microscopic features at high energy
and the macroscopic low-energy properties which are of
interest for the physics of condensed matter, and has enjoyed a
resurgence of interest recently\cite{polch,whiterg2,hewson}.

Our purpose here is to extend the RG method
to the study of {\it finite} fermi systems.
In particular, we introduce a new numerical RG technique
which, motivated by the case of
$C_{60}$, we apply to the study of one-dimensional rings
and a truncated tetrahedral lattice, as a prelude to
the study of the full $C_{60}$\   problem.
We note that the character of the problem is now
different from the case of infinite systems.
We can no longer rely on the flow to a fixed point.
However, as long as the
problem initially has a separation of energy scales,
the renormalization group is still useful for obtaining low
energy behavior,
with a number of advantages over other methods.
As described below,
our results are valid for a range of
coupling strengths considerably
beyond the reach of
perturbation theory.
The RG resums
the perturbation series to infinite order and
does not have the convergence problems of perturbation theory
as the system size increases.
Unlike selective resummations of perturbation theory,
the RG naturally yields a
criterion for deciding when it has reached the limit of
its validity.
We find that our method can reproduce
important qualitative
features missed by traditional many-body approaches.
We are also not limited by system size as severely as
numerical exact diagonalization methods, since the RG is a polynomial
time algorithm.
Finally we note that the phenomenon of interest here, namely
pair-binding (see below),
cannot occur in the infinite volume limit
(as it would correspond to a negative compressibility),
and hence is accessible only to methods
where the finite size of the system is explicitly taken
into account. The immediate precursors of this method are the basis set
reduction techniques\cite{iwata,brandow} used in atomic and nuclear systems.
This method holds promise
for the systematic investigation of a large variety of other problems,
as we discuss at the end of the paper.

We focus our attention here on the effective interaction between
particles at the fermi surface
$U^*$ (or equivalently the pairing energy)
\begin{equation}
 U^* = E_{0}(N) - 2 E_{0}(N-1) + E_{0}(N-2) ,
\label{mu}
\end{equation}
where $E_{0}(N)$ is the ground state energy for $N$ electrons.
A particularly intriguing proposal, which has been put forward to
explain  the superconductivity of $C_{60}$, is
that the repulsive bare interaction between electrons
can give rise to pair-binding ($U^*<0$) on a single
buckyball\cite{bask,chak}.
A number of papers\cite{fye,white,assa,goff}
have investigated this possibility and
demonstrated its existence in simple models.
We revisit this issue here
for several reasons.
The authors of Ref. \cite{assa}
in general found pair-binding
only at the limits of validity of their
second-order perturbation theory.
These and other workers\cite{goff}
also found
that pair-binding (in second-order perturbation theory)
is very sensitive to features such as the range
of the interaction, and it would useful to put these results on a
more solid footing.
Exact diagonalization methods, while
corroborating and extending these results for relatively small
systems\cite{white}, are unable to reach the full
$C_{60}$\  problem in the foreseeable future.
The desire to overcome these limitations provided
one of the initial motivations for introducing the RG.
The availability of exact results also
makes this problem a valuable testing ground for
our method, especially since
the pair-binding energy is
a {\it difference} in bulk energies and hence a stringent
test of any approximation.

Our method is a generalization of Wilson's momentum-shell RG\cite{wils}.
For concreteness,
we consider an extended Hubbard model
of electrons hopping on a one-dimensional ring or
a truncated tetrahedral lattice,
\begin{equation}
H = -t\sum_{\langle ij,s\rangle}(c^\dagger_{is}c_{js}+h.c.)
+U\sum_{i} n_{i\uparrow}n_{i\downarrow}+{V}\sum_{\langle
ij\rangle} n_i n_j
\label{H}
\end{equation}
with a nearest neighbor hopping $t$ (= 1 in what follows),
an onsite interaction $U$, and a
nearest neighbor interaction $V$.
In the single particle basis which diagonalizes the quadratic term,
the path-integral representation
for the partition function $Z$ is then
\begin{equation}
 Z = \int D{\bar{\eta}} D{\eta} \exp{S}
\label{Z}
\end{equation}
where the action S (at zero temperature) is of the general form
\widetext
\begin{eqnarray}
S &=& \sum_{ks}\int_{-\infty}^{\infty}{\bar{\eta}_{ks}}{\cal Z}(k)
(i\omega-\epsilon(k))\eta_{ks}(\omega) \nonumber \\
&-&\sum_{\{k_i\},ss'}
\int_{-\infty}^{\infty} {d\omega_1\cdots d\omega_4\over(2\pi)^3}
\delta(\omega_1+\omega_2-\omega_3-\omega_4)
V(\{k_i\}){\bar{\eta}}_{k_1s}
{\bar{\eta}}_{k_2s'}\eta_{k_3s'}\eta_{k_4s}
\label{S}
\end{eqnarray}
\narrowtext
The Grassman variables $\bar{\eta}, \eta$
represent electron creation
and annihilation operators.
Initially the wavefunction renormalization is ${\cal Z}(k)=1$, and
$\epsilon(k)$ and $V(\{k\},\omega)= V_{0}(\{k\})$ have their bare values.
\begin{eqnarray}
\epsilon(k)&=&-2\cos(k)\nonumber \\
V_0(k_1,k_2,k_3,k_4)&=&\delta(k_1+k_2-k_3-k_4){U+2V\cos(k_1-k_4)\over
N}
\label{barev}
\end{eqnarray}
Note that the matrix element
 is reduced by a factor of N (where N is the number of
lattice sites) from its real-space value.
We also add to the model a parameter
$\alpha$, having a value between 0 and 1, which
multiplies all umklapp
matrix elements.
This allows us to describe in a heuristic way
the transition from a spatial continuum (with no umklapp processes)
to the above tight-binding
lattice. (This is a reasonable approximation
for states away from the Brillouin zone boundary).

Since we are interested in the low-energy, long wavelength
behavior of the system, we
systematically integrate out shells of high-energy variables
(states lying far from the fermi surface)
one by one in the partition function, using
standard diagrammatic perturbation theory.
This procedure yields an effective
action of the same form for the remaining shells, but with modified values of
${\cal Z},\epsilon$, and $V$. The modified propogators and interaction
matrix elements are then used in the next integration step.
We continue renormalizing
until we are left with only
the states at the Fermi level.
We then convert the effective action
to Hamiltonian form and diagonalize it exactly.
If some of the couplings
grow larger than the energy denominators in the
perturbative analysis, we have to stop the RG flow and resort to
 exact diagonalization in order to extract
information.
It is one of the strengths of our method that the
formalism itself
tells us when it is no longer valid.
Note that even in this case, the RG,
being a polynomial time algorithm, is of considerable help in reducing
the work of an exact diagonalization routine.

Now some technical comments.
We integrate out a shell by expanding to two loops
in all propagators and four-point couplings.
Since we keep the
effect of all higher energy levels
on the level presently being renormalized,
this procedure corresponds to keeping
the effect of all six and eight point couplings on the four fermi
interaction.
There are several reasons why we expand to two-loop order.
{}From a theoretical standpoint,
this is required in order to get
sensible results for one-dimensional fermi systems
in the infinite volume limit \cite{solyom}.
This is  quite a general statement, since, as
shown by Shankar\cite{shankar},
{\it all} fermi systems are equivalent to an infinite set of coupled
one-dimensional fermi systems.
We also do not expect to have to go beyond two loops.
 In fact, one criterion for the
breakdown of the RG is the appearance of significant
three-loop corrections, for then we expect all higher
loop corrections to also be important.
Finally, we find
that by going to two loops, we can reproduce
certain qualitative features of the exact solution,
as detailed below.

After integrating out a shell,
we expand the quadratic term in the action
to first order in frequency and  extract the
the corrected ${\cal{Z}}$ factor.
We  drop the frequency dependence of the interaction, since we
are interested in low-frequency behavior (frequencies smaller than the
energy gaps in the system).
Other than these simplifications, we keep {\it all} the
couplings $V(\{k\})$ and
 do not prejudice the formalism as to which couplings are most
important.

Finally, to convert to hamiltonian form the Grassman variables must be
rescaled by $\sqrt{\cal{Z}}$, thus rescaling the interaction by ${\cal{Z}}^2$.
It is also necessary to
calculate a composite operator renormalization ${\cal Z}_2$, which
describes the propagation of two particles (as opposed to just one)
at the fermi surface. These and other details will be described elsewhere.

We now turn to our results.
Fig.~\ref{ring8} and Fig.~\ref{ring16}
compare the results for $U^*$ using second and third order
perturbation theory,
two-loop RG, and exact diagonalization,
for an eight and sixteen site ring (exact numbers for the 16-site case
taken from
\cite{fye}) respectively.
The RG is valid over a considerably larger range of $U$
than perturbation theory.  It starts
deviating from the exact result roughly
when we expect it to, namely when
\begin{equation}
{\frac{V}{Z\s^2 \Delta\epsilon}} \sim 1 ,
\label{dev}
\end{equation}
where $\Delta\epsilon$ is the smallest energy denominator (usually the
bandwidth)  prior to the last
step of the RG.
The RG is similarly successful in
the case of a
truncated tetrahedron, as demonstrated in
Fig.~\ref{tt}, where the exact results are those
of  White {\it et al}\cite{white}.

Several authors\cite{white,khleb,gunnar}
have noted
that $U^*$ can
exhibit an unusual non-monotonic
dependence on $U$, as seen in Fig.~\ref{ring8}.
While this behavior is correctly given by the RG,
it cannot be reproduced by the method traditionally used in
the theory of superconductivity, i.e. summing ladder diagrams\cite{morel}.
This latter approach yields a monotonic dependence
of $\mu^*$ on $\mu$,
\begin{equation}
\mu^* = {\mu\over 1+\mu \log(W/\omega_d)}
\label{mulad}
\end{equation}
where the dimensionless quantities $\mu$, $\mu^*$
are related by a density of states factor to
$U$, $U^*$. Thus,  selective resummations of
perturbation theory can miss important qualitative features
of the problem.

The above non-monotonic behavior implies that
a minimum in $U^*$ can be achieved by
tuning $U$ (actually $U/t$).
Thus, even if purely electronic interactions are not
responsible for the superconductivity in $C_{60}$,
attempts to tune $U/t$ (where we note that $t$ refers to
{\it intraball} and not interball hopping)
can be important for optimizing $T_c$.

We can gain insight into the
physical basis for this behavior by
investigating how the couplings transform under the
RG.  As expected,
a relatively small set of couplings (the forward scattering and umklapp
scattering of pairs at the same $k$) are growing\cite{rokh1},
while the others are disappearing.
By introducing variational wavefunctions which
minimize the energy with respect to the largest couplings,
we can discover in an unbiased manner which
correlations are most important in the system.
This is work in progress which will be reported elsewhere.

Fig.~\ref{phase} is a RG phase diagram for pair-binding
as a function of $V/U$ and $\alpha$ for the 12-site ring at $U=4$.
Pair-binding is quite sensitive
to the range of the interaction, as suggested by
earlier perturbation theory results\cite{assa,goff}, and also to
the  strength of the lattice. It is interesting to note that even for the
tight binding lattice the maximum $V/U$ for pair binding is $0.27$,
which approximately corresponds to a screening length of one intraring
lattice spacing (compare Ref.\cite{goff}).

To summarize, we have presented a numerical RG method that is a viable means of
obtaining information about the low-energy properties
of finite Fermi systems.
The RG is a systematic approach which transcends
a number of limitations of
perturbation theory, selective resummations,
and exact diagonalizations,
providing information not only about energies but also
particle correlations.
Future work in this area includes:
1) Going to larger systems (ultimately $C_{60}$), both through RG alone and by
   combining RG with exact diagonalization;
2) Exploring correlations by building
   variational wavefunctions inspired by RG;
3) Using finite-size scaling to obtain information about infinite systems;
4) Extending the formalism to investigate low-frequency dynamic properties,
   including susceptibilities;
5) Exploring properties at finite temperatures.

An important extension for the $C_{60}$\  problem will be to
include phonons in the formalism.  For the $C_{60}$\  crystal,
   the RG becomes a continuous process over a number of electronic
   bands.  Since the phonon energies are
   of the order of the conduction level bandwidth, the
   use of Migdal's approximation\cite{migdal} is not justified.
   The RG can systematically deal with this situation.
   The high-energy electronic shells will now
   renormalize the electrons and phonons simultaneously,
   modifying both the energies of the phonons and
   electron-phonon coupling in the conduction band.
   As before, we expect a few couplings to grow the fastest,
   reflecting the important correlations (e.g. superconducting)
   which are developing in the system, and
   these can then be treated by mean-field theory. Crystal screening
effects\cite{gunnar}
 add another layer of complication to any treatment of the real
problem.

Although we have concentrated on simple fullerene analogs, the technique
presented here may be fruitfully employed  in a number of problems.
 Potential
applications of the RG include investigating persistent currents
in metallic rings\cite{persistent}, correlations in Quantum
Dots\cite{qdots},
and the possible breakdown of
Fermi liquid theory in the thermodynamic limit\cite{mflt}.  In the latter case,
the behavior of the wavefunction renormalization ${\cal{Z}}$ will play a
prominent role.

\section*{Acknowledgments}
We gratefully acknowledge useful conversations with Professors A.Auerbach,
S.Chakravarty, S.Kivelson, D.Rokhsar, and R.Shankar,
the hospitality of the Aspen Center for Physics, where
some of this work was done, and the Boston University CM5. We also thank
Prof S.R.White for sending us the exact diagonalization data for the
truncated tetrahedron.

\begin{figure}
\caption{This figure shows the pair binding energy of an 8-site ring
with a nearest-neighbour interaction 7\% the on-site interaction. The
lattice is tight-binding ($\alpha=1$). The RG  approximates the
exact solution much better than either second- or third-order
perturbation theory for a wider range of $U$.}
\label{ring8}
\end{figure}

\begin{figure}
\caption{The same comparison as Figure 1. for a 16-site tight-binding Hubbard
ring
($V=0$, $\alpha=1.$). Once again the RG approximates the exact solution
much better than perturbation theory for a wide range of $U$.}
\label{ring16}
\end{figure}

\begin{figure}
\caption{The same comparison as Figures 1. and 2.
for the truncated tetrahedron with pure
Hubbard interactions. Once again the RG is the best approximation.}
\label{tt}
\end{figure}

\begin{figure}
\caption{The phase diagram for the 12-site ring at $U=4$ from RG.
Nearest neighbour interactions corresponding to a screening length of
one intraring lattice spacing or a slight weakening of the lattice can
destroy pair binding. }
\label{phase}
\end{figure}

\end{document}